\documentclass[]{aa}

\usepackage{graphicx,natbib}

\def\jh{\mbox{$\rm (J-H)$}}

\def\jk{\mbox{$\rm (J-K_s)$}}

\def\ebv{\mbox{$\rm E(B-V)$}}
\def\ejh{\mbox{$\rm E(J-H)$}}
\def\rc{\mbox{$\rm R_{core}$}}
\def\rl{\mbox{$\rm R_{lim}$}}
\def\rt{\mbox{$\rm R_{tidal}$}}
\def\ms{\mbox{$\rm M_\odot$}}
\def\ds{\mbox{$\rm d_\odot$}}

\def\feh{\mbox{$\rm [Fe/H]$}}
\def\jj{\mbox{$\rm J$}}
\def\hh{\mbox{$\rm H$}}
\def\ks{\mbox{$\rm K_s$}}

\begin{document}

\title{FSR\,584 - a new globular cluster in the Galaxy?}

\author{E. Bica\inst{1} \and C. Bonatto\inst{1} \and S. Ortolani\inst{2} \and B. Barbuy\inst{3}}
\offprints{Ch. Bonatto}

\institute{Universidade Federal do Rio Grande do Sul, Departamento de Astronomia\\
CP\,15051, RS, Porto Alegre 91501-970, Brazil\\
\email{charles@if.ufrgs.br, bica@if.ufrgs.br}
\mail{charles@if.ufrgs.br} 
\and
Universit\`a di Padova, Dipartimento di Astronomia\\
Vicolo dell'Osservatorio 5, I-35122 Padova, Italy; \email{sergio.ortolani@unipd.it}
\and
Universidade de S\~ao Paulo, Departamento de Astronomia\\
Rua do Mat\~ao 1226 S\~ao Paulo 05508-900, Brazil; \email{barbuy@astro.iag.usp.br}}

\date{Received --; accepted --}

\abstract
{}
{We investigate the nature of the recently catalogued star cluster candidate FSR\,584, which
is projected in the direction of the molecular cloud W\,3 and may be the nearest globular
cluster to the Sun.}
{2MASS colour-magnitude diagrams, the stellar radial density profile, and proper motions are
employed to derive fundamental and structural parameters.}
{The colour-magnitude diagram morphology and the radial density profile show that FSR\,584 is 
an old star cluster.
With proper motions, the properties of FSR\,584 are consistent with a metal-poor ($\feh\approx-2$)
globular cluster with a well-defined turnoff and evidence of a blue horizontal-branch. FSR\,584 might
be a Palomar-like halo globular cluster that is moving towards the Galactic plane, currently at
$\approx20$\,pc above it. The distance from the Sun is $\ds\approx1.4$\,kpc, and it is located at
$\approx1$\,kpc outside the Solar circle. The radial density profile is characterized by a core
radius of $\rc=0.3\pm0.1$\,pc. However, we cannot exclude the possibility of an old open cluster. }
{Near-infrared photometry coupled to proper motions support the scenario where FSR\,584 is a new globular
cluster in the Galaxy. The absorption is $A_V=9.2\pm0.6$, which makes it a limiting object in the
optical and explains why it has so far been overlooked.}


\titlerunning{A new globular cluster in the Galaxy?}

\authorrunning{E. Bica et al.}

\maketitle

\section{Introduction}
\label{intro}

Since they are long-lived, globular clusters (GCs) formed in the initial phases of the Galaxy
may preserve information in their structure and spatial distribution that is essential for probing
the early Milky Way physical conditions. Derivation of the present-day spatial distribution of GCs, 
as well as their physical and chemical properties, is important to better understand the formation 
and evolution processes, and trace the geometry of the Galaxy (\citealt{MvdB05}; \citealt{GCProp}).

Over a Hubble time, the structure of GCs, especially the less-massive ones, is affected by external
processes such as the Galactic tidal stress and dynamical friction, and internal ones such as mass 
loss by stellar evolution, mass segregation and evaporation (e.g. \citealt{Khalisi07}; \citealt{Lamers05};
\citealt{GnOs97}). The net result of these processes is a decrease in the mass of the clusters that
may accelerate the core collapse phase for some (\citealt{DjMey94}, and references therein). 
The bottom line is that internal dynamical processes are affected by external ones, whose strength also depends 
on the GC Galactocentric distance (e.g. \citealt{DjMey94}; \citealt{vdBergh91}; \citealt{CheDj89}).

As remarked by \citet{vdBergh91}, the Galactic halo is a friendly environment to clusters of any size. 
Small GCs formed in the halo should have a high probability of surviving to date, and the scarcity of 
both small halo GCs and large ones close to the Galactic center could result from a physical correlation
established at cluster formation. Thus, the discovery and characterization of low-mass GCs plays an important 
role, both to improve the statistics at the faint-end of the GC-parameter distribution, and to set observational
constraints on the cluster-dynamical processes discussed above.

\begin{figure*}
\begin{minipage}[b]{0.50\linewidth}
\includegraphics[width=\textwidth]{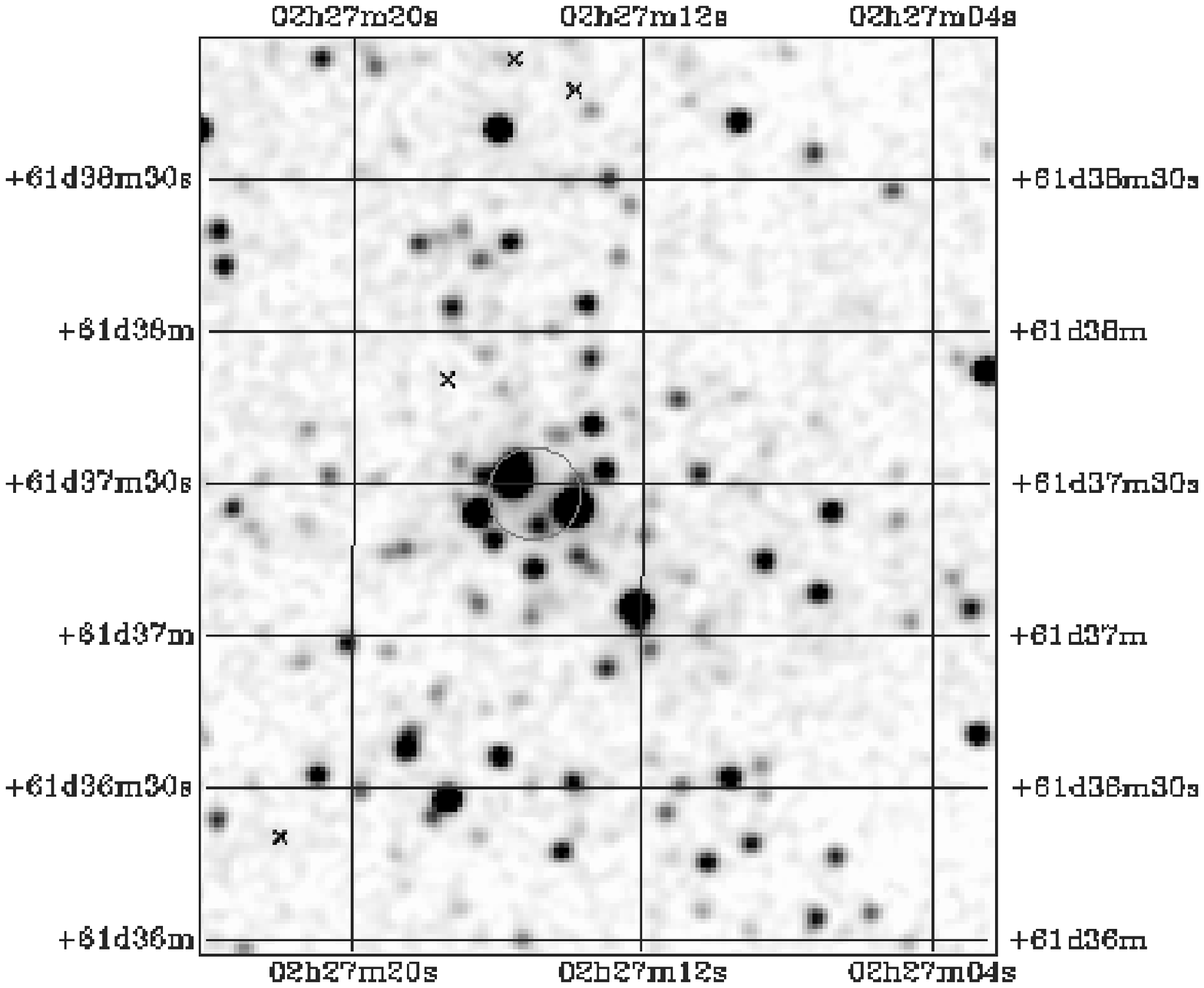}
\end{minipage}\hfill
\begin{minipage}[b]{0.50\linewidth}
\includegraphics[width=\textwidth]{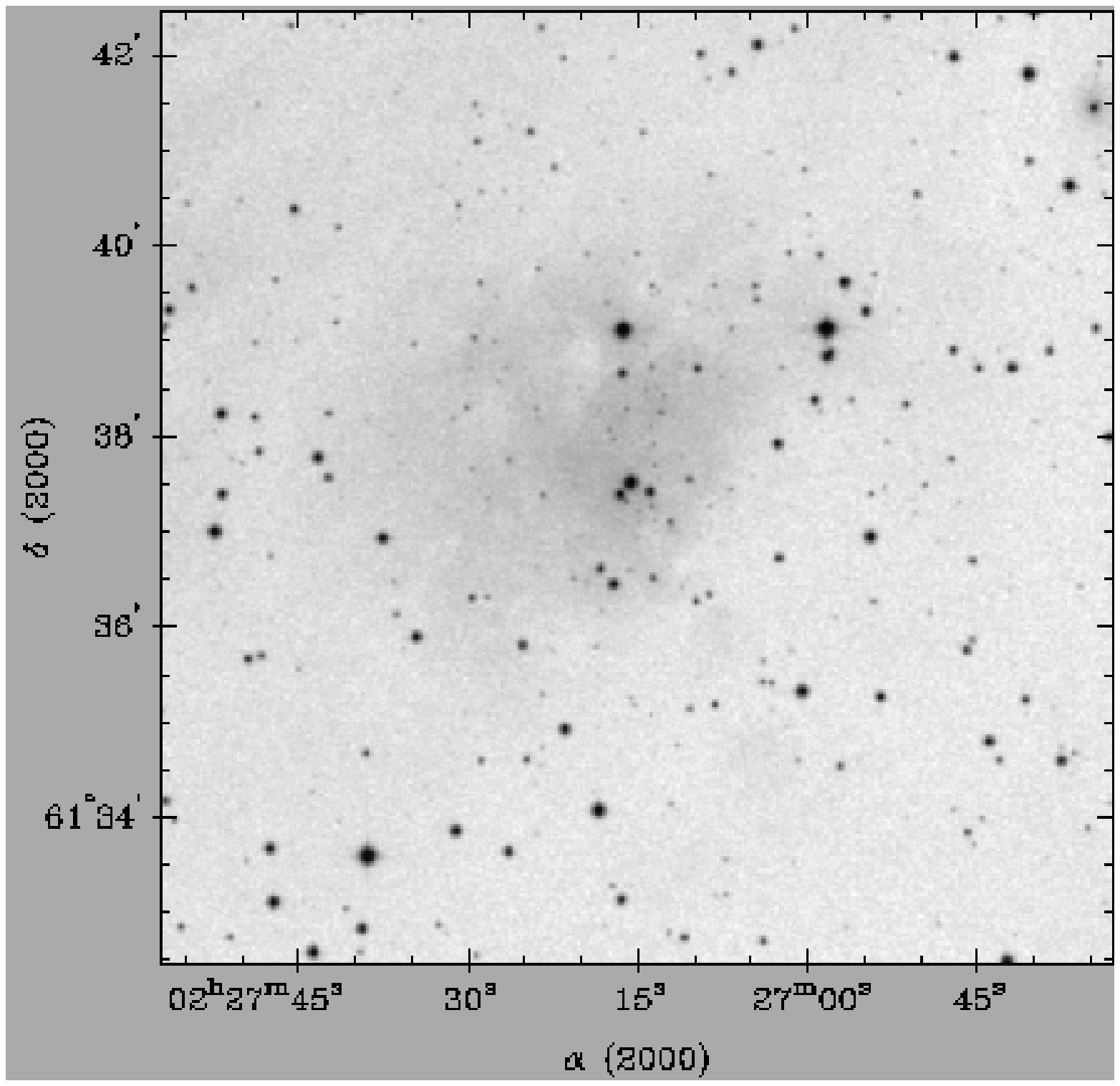}
\end{minipage}\hfill
\caption[]{Left: $3\arcmin\times3\arcmin$ 2MASS \hh\ image of FSR\,584. The small circle marks
the central region of the cluster (Sect.~\ref{PhotPar}). North is up and East is left. Right:
$10\arcmin\times10\arcmin$ XDSS R image, where an optical nebula of diameter of $\sim2.5\arcmin\approx1$\,pc
(Sect.~\ref{Disc}) is projected in the central cluster area.}
\label{fig1}
\end{figure*}

The number of known GCs in the Galaxy has been slowly increasing as deeper surveys have been carried out. The 
compilation of Harris (1996, and the 2003 update\footnote{\em http://physun.physics.mcmaster.ca/Globular.html})
contains 150 GCs. \citet{KMB2005} discovered the far-IR GC GLIMPSE-C01. \citet{Carraro05} identified Whiting\,1 
as a young halo GC. Two stellar systems detected with the Sloan Digital Sky Survey (SDSS) in the outer halo,
Willman\,1 (SDSS\,J1049$+$5103) and SDSS\,J1257$+$3419, might be GCs or dwarf galaxies (\citealt{Willman05};
\citealt{Sakamoto06}). \citet{OBB06} identified AL\,3 as a new GC in the bulge with a blue horizontal branch 
(HB). More recently, \citet{FMS07} found evidence that FSR\,1735 is a GC in the inner Galaxy, and
\citet{Belokurov07} found the faint halo GC SEGUE\,1 using SDSS. Finally, \citet{Koposov07} reported the 
discovery of two very-low luminosity halo GCs (Koposov\,1 and 2) detected with SDSS.

In a recent observational effort to uncover potential star clusters, \citet{FSR07} carried out
an automated search for stellar overdensities using the 2MASS\footnote{{\em
http://www.ipac.caltech.edu/2mass/releases/allsky/}} database for $b<20^\circ$, which resulted in a
list of 1021 candidates. Based on diagnostic diagrams involving number of stars, core radius and 
central density, 9 of these were classified as GC candidates, whilst the majority were
open cluster candidates. FSR\,584 is among the latter. A systematic inspection of the respective 2MASS 
images revealed that the candidate FSR\,584 is a relatively populous star 
cluster with a pronounced core. In fact, in the 2MASS atlas, the cluster image (Fig.~\ref{fig1}) resembles 
a Palomar GC (e.g. \citealt{ORS85}).

In the present work, we investigate the nature of FSR\,584. In Sect.~\ref{PhotPar} we analyze near-IR
colour-magnitude diagrams (CMDs), proper motions (PMs) and cluster structure. In Sect.~\ref{Disc} we discuss
cluster properties. Concluding remarks are given in Sect.~\ref{Conclu}.

\section{Photometric parameters}
\label{PhotPar}

The original coordinates of FSR\,584 (\citealt{FSR07}) are slightly shifted with respect to the
central concentration of stars as seen in 2MASS images (Fig.~\ref{fig1}). The revised values
are $\alpha(J2000)=02^h\,27^m\,15^s$ and $\delta(J2000)=61^\circ\,37\arcmin\,28\arcsec$, which
correspond to the Galactic coordinates $\ell=134.06^\circ$ and $b=+0.84^\circ$. We are dealing with
a $\rm2^{nd}$ quadrant cluster projected very close to the plane in Cassiopeia. The following analysis
is based on 2MASS photometry and tools as described in \citet{BB07}.

$\jj\times\jh$ CMDs of different extractions around the cluster center are presented in
Fig.~\ref{fig2}. Panel (a) shows the region $R<3\arcmin$ that contains most of the cluster stars
(Sect.~\ref{Struc}). A disk-like population appears to merge into a redder component at $\jh>1.1$.
Panel (b) contains an equal area annular extraction at $R\approx20\arcmin$. The disk component stands out as
in (a) but the redder one is vanishing. The red component is essentially absent at $R\sim30\arcmin$ (panel c).

\subsection{Field-star decontamination}
\label{Decont}

To disentangle the intrinsic CMD morphology from the field, we apply the decontamination algorithm 
described in \cite{BB07}. The algorithm works on a statistical basis by measuring the relative
number-densities of candidate field and cluster stars in small cubic CMD cells whose axes correspond to 
the magnitude \jj\ and the colours \jh\ and \jk\ (considering as well the $1\sigma$ uncertainties in the 
2MASS bands). These are the 2MASS colours that provide the maximum variance among CMD sequences for star
clusters of different ages (e.g. \citealt{TheoretIsoc}).

Basically, the algorithm {\em (i)} divides the full range of magnitude and colours of a given CMD into a 
3D grid, {\em (ii)} computes the expected number-density of field stars in each cell based on the number 
of comparison field stars with magnitude and colours compatible with those in the cell, and {\em (iii)} 
subtracts the expected number of field stars from each cell. The algorithm results are sensitive to local 
variations of field-star contamination with colour and magnitude (\citealt{BB07}). Cell dimensions are
$\Delta\jj=0.5$, and $\Delta\jh=\Delta\jk=0.25$, which are large enough to allow sufficient star-count 
statistics in individual cells and small enough to preserve the morphology of the CMD evolutionary 
sequences. As a comparison field, we use the region $20<R(\arcmin)<30$ around the cluster center to obtain
representative background statistics.

Three different grid specifications in each dimension are used to minimize potential 
artifacts introduced by the choice of parameters. For instance, for a CMD grid starting at magnitude
$J_o$ (and cell width $\Delta\jj$), additional runs for $J_o\pm\frac{1}{3}\Delta\jj$ are included. 
Considering the 2 colours as well, 27 different outputs are obtained, from which the average number 
of probable cluster stars $\langle N_{cl}\rangle$ results. Typical standard deviations of 
$\langle N_{cl}\rangle$ are at the $\approx2.5\%$ level. The final field star-decontaminated CMD contains 
the $\langle N_{cl}\rangle$ stars with the highest number-frequencies. Stars that remain in the CMD after 
the decontamination are in cells where the stellar density presents a clear excess over the 
field. Consequently, they have a significant probability of being cluster members. Further details on the
algorithm, including discussions on subtraction efficiency and limitations, are given in \citet{BB07}.

The resulting field star decontaminated CMD derived from the extraction $R<3\arcmin$ (panel a) is shown 
in panel (d). As expected, most of the disk component is removed, while a populous red component shows 
up. This extraction, which corresponds to the bulk of the cluster stars (see Fig.~\ref{fig5}), indicates
that we are dealing with a relatively populous star cluster, as suggested by the prominent main sequence 
(MS) at $1.0\la\jh\la2.0$. Comparable results are derived with the additional 2MASS filter \ks, by means 
of $\jj\times\jk$ CMDs (Fig.~\ref{fig3}).

\subsection{The central CMD and proper motions}
\label{CentralCMD}

To minimize field contamination issues, we examine in Fig.~\ref{fig3} the observed CMD of a central
($R<1\arcmin$) extraction. A similar procedure was fundamental in the analysis that unveiled AL\,3 as 
a GC (\citealt{OBB06}). The central stellar density is so high with respect to the background that when 
applied to this CMD, the field decontamination would remove only 2 stars (Fig.~\ref{fig3}). Besides the 
well-defined MS and turnoff (TO) sequences indicating old populations, the CMD displays a 
group of blue stars suggesting a blue HB. To test the possibility that FSR\,584 is an old open cluster (OC),
we compared its CMD morphology with that of the $\approx7$\,Gyr OC NGC\,188 (\citealt{NGC188}). The locus 
occupied by the probable HB stars in FSR\,584 is very different from that of the blue-stragglers in NGC\,188.
Although hot HB candidates have been detected in old open clusters (\citealt{Landsman1998}), they are 
rare, whereas blue HB stars are typical of metal poor GCs. On the other hand, the CMD morphology of the 
$R<1\arcmin$ region of FSR\,584 is similar to that of the GC NGC\,6397 with metallicity $\feh=-1.95$
(\citealt{Harris96}), built with 2MASS photometry of the $R<3\arcmin$ region. As a reference for the MS, 
TO and red giant branch (RGB) sequences, we use a 10\,Gyr Padova (\citealt{Girardi02}) isochrone of 
$\feh=-2.0$. Young clusters, as a rule, have less densely populated TOs (\citealt{N4755}) and do not 
account for the blue HB stars. The results are consistent with a metal-poor GC, containing some blue HB 
stars. FSR\,584 appears to be similar to the low-mass GCs AL\,3, Palomar\,13 (\citealt{Siegel01}), and AM\,4
(\citealt{InmanC87}), which contain about 10 giants.

\begin{figure}
\resizebox{\hsize}{!}{\includegraphics{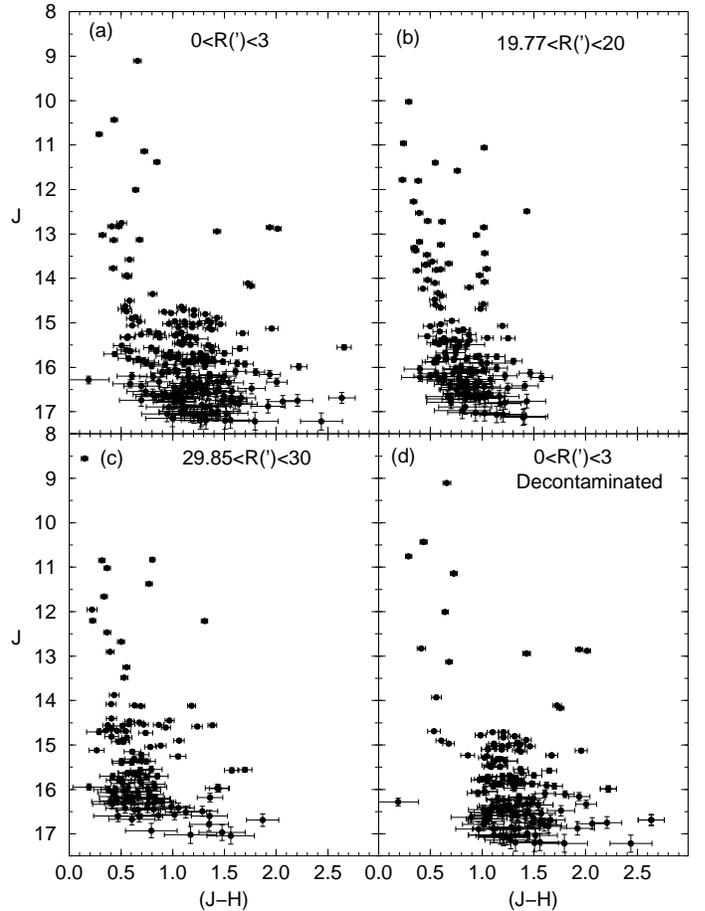}}
\caption{Equal area CMDs extracted as a function of distance from the cluster center, as indicated in 
the panels. Photometric errors are shown. From (a) to (c) the disk component remains, while the red star 
content vanishes. Panel (d): field decontaminated CMD of panel (a).}
\label{fig2}
\end{figure}

\begin{figure}
\resizebox{\hsize}{!}{\includegraphics{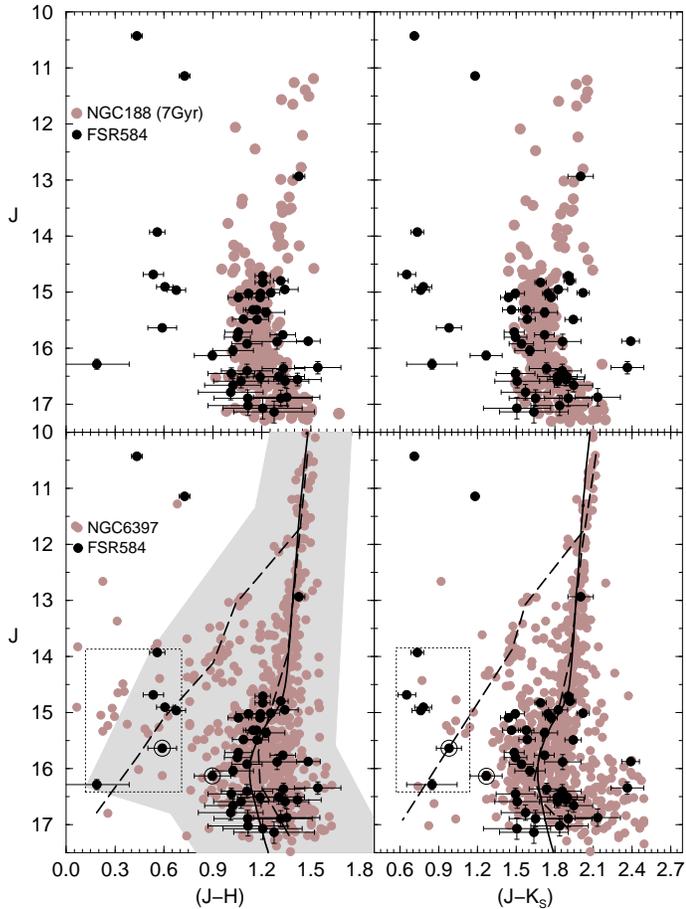}}
\caption{Top: $R<1\arcmin$ CMD of FSR\,584 (black circles) compared to a CMD of the 7\,Gyr open cluster 
NGC\,188 (gray). Bottom: comparison with the $R<3\arcmin$ CMD of the GC NGC\,6397 (gray) and corresponding 
mean locus (dashed line). Solid line: 10\,Gyr Padova isochrone with $\feh=-2.0$. Dotted rectangle: locus of 
probable HB stars in FSR\,584. Shaded area: colour-magnitude filter used to build the stellar radial 
density profile. Field decontamination would remove the 2 encircled stars.}
\label{fig3}
\end{figure}

\begin{figure}
\resizebox{\hsize}{!}{\includegraphics{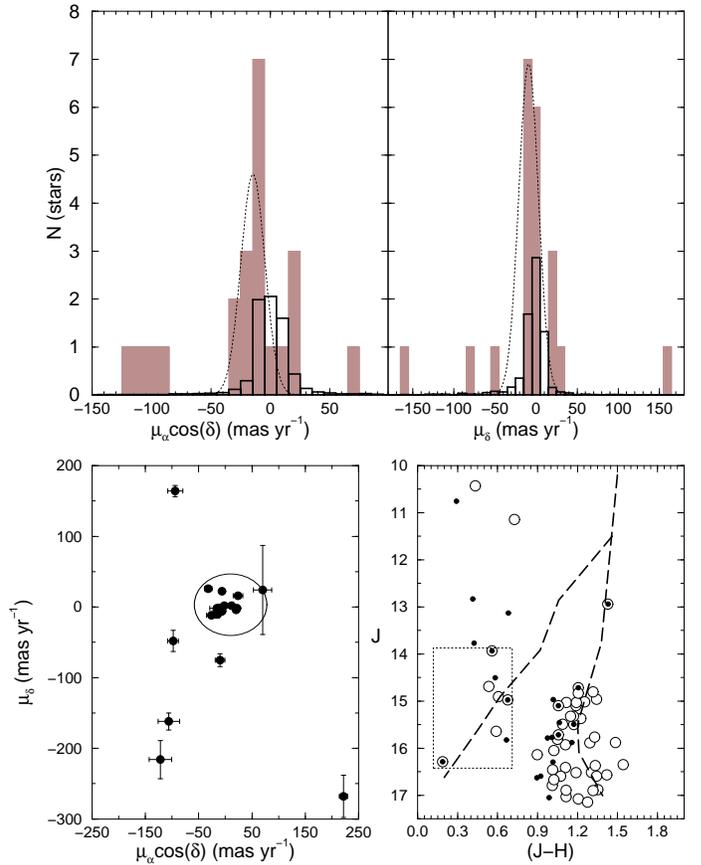}}
\caption{Top: Proper motions of the stars within $R<2\arcmin$ (shaded histogram) compared to those at 
$40\arcmin<R<50\arcmin$ (empty), corrected for the different solid angles. Dotted line: Gaussian fit to 
the PM distribution of the central region. Bottom-left: PM components of the $R<2\arcmin$ stars of FSR\,584. 
Those located in the clump (circle) occur mostly in the MS and HB of FSR\,584 (circles with dot - right 
panel). Dashed line: fiducial line of NGC\,6397 (from Fig.~\ref{fig3}).}
\label{fig4}
\end{figure}

Isochrone fits are independently applied to CMDs in both colours to derive photometric parameters 
(Fig.~\ref{fig3}). The fits imply a reddening $\ejh=0.93\pm0.05$, which converts to $\ebv=3.0\pm0.2$, and
$A_V=9.2\pm0.6$\footnote{Reddening transformations are $A_J/A_V=0.276$, $A_H/A_V=0.176$, $A_{K_S}/A_V=0.118$,
and $A_J=2.76\times\ejh$ (\citealt{DSB2002}), with $R_V=3.1$.}. The distance of FSR\,584 from
the Sun is $\ds=1.4\pm0.2$\,kpc, virtually crossing the plane at a height of $\approx20$\,pc. Adopting
$R_\odot=7.2$\,kpc as the Sun's distance to the Galactic center (\citealt{GCProp}), we find that FSR\,584 
is located $\approx1$\,kpc beyond the Solar circle. Among the optical GCs, M\,4 and NGC\,6397 are 
located at $\ds=2.2$ and $\ds=2.3$\,kpc, respectively (\citealt{Harris96}). The IR GC GLIMPSE-C01, is also 
quite close at $\ds\approx4$\,kpc (\citealt{KMB2005}). This implies that FSR\,584 is the nearest GC.

We further tested the nature of FSR\,584 with proper motion data from
NOMAD\footnote{\em http://vizier.u-strasbg.fr/viz-bin/VizieR?-source=I/297}. Since the correspondence between
2MASS and NOMAD is not complete, we extracted the PM of the stars within $R<2\arcmin$ of FSR\,584 to increase
statistics. Besides, the stellar density profile of FSR\,584 for $R<2\arcmin$ presents a high contrast with 
respect to the background (Fig.~\ref{fig5}). The PM distributions of these stars present differences both in
the number of stars per PM bin and average values, especially in right ascension, with respect to the 
distant comparison ring $40\arcmin<R<50\arcmin$ (Fig.~\ref{fig4}). The average 
values for the central region are $\rm\langle\mu_\delta\rangle=-9.1\pm1.4\,mas\,yr^{-1}$ and
$\rm\langle\mu_\alpha\cos(\delta)\rangle=-14.7\pm3.3\,mas\,yr^{-1}$. Most of the $R<2\arcmin$ stars populate a 
clump in the PM plane, $\rm|\mu_\delta|=|\mu_\alpha\cos(\delta)|\la35\,mas\,yr^{-1}$, as expected for a star 
cluster. In the $R<1\arcmin$ observed CMD of FSR\,584, these stars occur mostly in the MS/TO and HB sequences
(Fig.~\ref{fig4}). Although in smaller number because PM data are from the optical, red stars essentially share 
the same PM as the blue stars. This motion in common supports the scenario where the bluer stars (HB) belong to 
the cluster, which is compatible with FSR\,584 being a metal-poor halo GC. Assuming the average PM values of 
these stars as representative of FSR\,584, we estimate that it is moving towards the Galactic plane at an angle 
of $\approx71\pm5^\circ$, consistent with the motion of a halo object. In the case of an old open
cluster, it would have very peculiar kinematics.

\subsection{Structural parameters of FSR\,584}
\label{Struc}

\begin{figure}
\resizebox{\hsize}{!}{\includegraphics{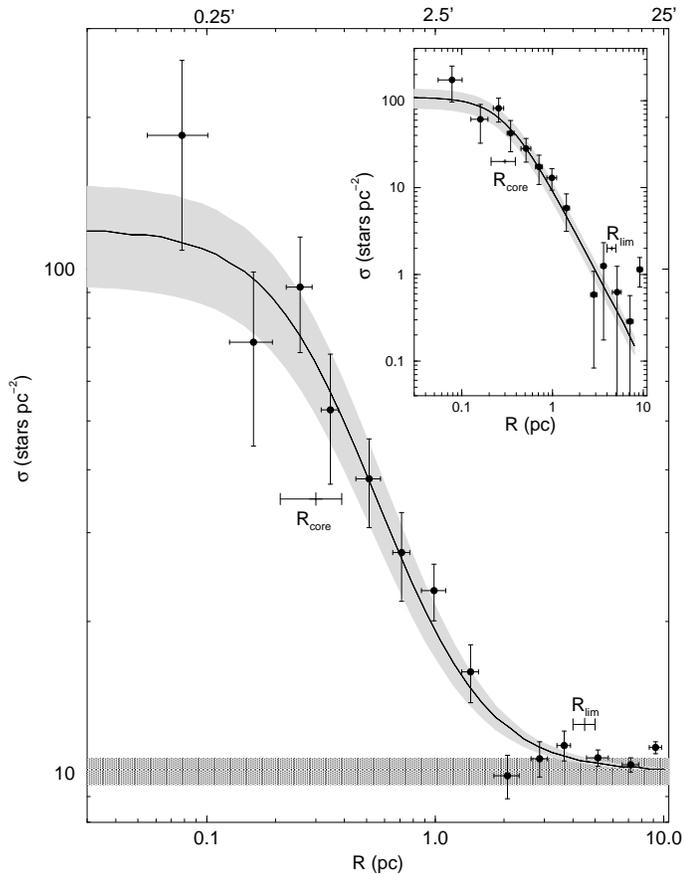}}
\caption{Stellar RDP (filled circles) of FSR\,584 in absolute scale. Solid lines: best-fit King
profile. Horizontal shaded region: offset field stellar background level. Core and limiting
radii are indicated. Gray regions: $1\sigma$ King fit uncertainty. The angular scale is in the upper 
abscissa. Inset: background-subtracted RDP together with King fit.}
\label{fig5}
\end{figure}

The structure is investigated using the stellar radial density profile (RDP), built with colour-magnitude 
filtered photometry for $R<30\arcmin$ (\citealt{BB07}). The filter (Fig.~\ref{fig3}) removes contamination 
of stars with colours deviant from cluster sequences in the CMD. The resulting profile (Fig.~\ref{fig5})
presents a prominent excess over the background especially for $R<4$\,pc. Most of the RDP follows
King's (1966) law. However, the innermost point has a $1\sigma$ excess density over the King fit, which 
might suggest post-core collapse. GCs with that structure occur mostly in the bulge, but some halo GCs like 
NGC\,6397 also have it (\citealt{CheDj89}; \citealt{TKD95}).

Despite the central excess, the King fit provides a core radius of $\rc=0.3\pm0.1$\,pc ($\approx0.75\arcmin$).
The cluster limiting radius (the distance from the cluster center where RDP and background statistically
merge) is $\rl=4.5\pm0.5$\,pc ($\approx11\arcmin$). Within uncertainties, the present value of \rc\ is
compatible with that in \citet{FSR07}, $\rc=1\arcmin$. They estimate a tidal radius of $\rt=50.6\arcmin$,
about 5 times larger than \rl. However, dust, especially in the W\,3 molecular cloud (Sect.~\ref{Disc}),
introduces dips in radial star counts for $R>30\arcmin$ that precludes derivation of \rt\ from the King
fit. The inset of Fig.~\ref{fig5} displays the RDP of FSR\,584 in the usual, background-subtracted presentation
for GCs.

\section{Discussion}
\label{Disc}

An unusual property of FSR\,584 is that it is projected in the direction of the large
($\rm diameter\approx50\arcmin$), nearby molecular cloud/star-forming region W\,3 (\citealt{Carpenter00}). 
Within uncertainties, the distance of FSR\,584 is compatible with $\ds\approx1.88$\,kpc
(\citealt{Kharchenko05}) for the cluster IC\,1805 in the neighboring cloud W\,5, at a comparable distance
(\citealt{Carpenter00}). 

\begin{figure}
\resizebox{\hsize}{!}{\includegraphics{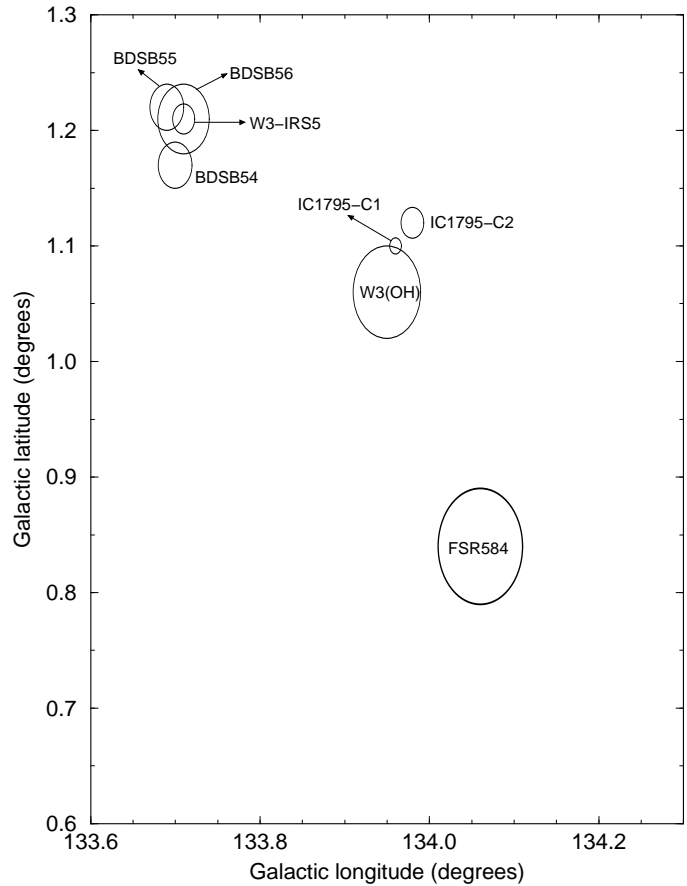}}
\caption{Angular distribution and size of embedded clusters in the W\,3 cloud together with
those of FSR\,584. The 3\arcmin radius of FSR\,584 in this figure corresponds basically to
the region containing the bulk of cluster stars (Fig.~\ref{fig5}), thus encompassing the
cluster region shown in the 2MASS image (Fig.~\ref{fig1}). Orientation as in Fig.~\ref{fig1}.}
\label{fig6}
\end{figure}

Angular positions and sizes of the known embedded clusters in W\,3 are compared with those of
FSR\,584 in Fig.~\ref{fig6}. The embedded clusters are from \citet{Carpenter00}, \citet{PAP018},
\citet{REF009}, \citet{REF010}, and \citet{PAP023}. FSR\,584 is spatially separated from the
known embedded clusters in W\,3.

An optical nebula is present in blue and red XDSS and DSS images (Fig.~\ref{fig1}). This reflection and/or
ionizing nebula might be associated with the cluster. The latter peculiar relation with an old cluster is possible
since blue HB stars normally attain $10^4$\,K, and hot ones may reach $2\times10^4$\,K (\citealt{MHB1977}).
Most of the nebula is contained within $R\approx2.5\arcmin$ ($\approx1$\,pc). Following photo-dissociation,
$10^4$\,K stars can ionize gas in a comparable volume for densities of $\rm\sim1~H~atom~\,cm^{-3}$, while
$2\times10^4$\,K stars may ionize gas as far as $8-10$\,pc.

FSR\,584 appears to be a halo GC moving towards the plane (Sect.~\ref{CentralCMD}). Besides disk-shocking, 
it might be suffering additional tidal stress due to its proximity to the massive W\,3, W\,4 and W\,5 
molecular cloud complex (\citealt{Wielen91}). W\,3 itself is relatively massive with $\approx10^5\ms$
(\citealt{Lada87}). In the long term, dynamical heating of a GC is expected as a consequence of tidal 
interactions by shocks due to disk and bulge crossings, as well as encounters with massive molecular clouds. 
This enhances the rate of low-mass star evaporation and accelerates the process of core collapse 
(\citealt{DjMey94}). A combination of these effects could enhance the cluster dynamical evolution (\citealt{BB07},
and references therein) with core contraction and halo expansion (e.g. \citealt{GoBa06}). This might explain a
central density excess over the King profile (Fig.~\ref{fig5}) and, at the same time, account for the presence
of cluster stars driven to distances of up to $\sim20\arcmin$\ ($\sim8$\,pc) from the cluster center
(Fig.~\ref{fig2}). A similar effect was detected in the intermediate-age open cluster BH\,63 (\citealt{FaintOCs}), 
which shows a post-core collapse structure, probably because of interaction with a neighboring dust cloud.

\section{Concluding remarks}
\label{Conclu}

FSR\,584 was found by \citet{FSR07} in an automated star cluster survey using 2MASS. The near-IR CMD
morphology (Sect.~\ref{CentralCMD}) combined to the stellar radial density profile (Sect.~\ref{Struc})
provide convincing evidence that FSR\,584 is an old star cluster. Together with these results, 
PMs compatible with the halo suggest an as yet unidentified GC in the Galaxy. Deeper IR 
photometry complemented with spectroscopy are necessary to settle the nature of FSR\,584, as a 
GC (old or young) or an old open cluster with peculiar kinematics. We point out that young 
GCs such as Pal\,1 (\citealt{Rosenberg98}) 
and Whiting\,1 (\citealt{Carraro05}) are low-mass objects with hardly any giants and essentially no HB 
stars. This is also the case for the recently discovered low-mass halo GCs Koposov\,1 and 2 
(\citealt{Koposov07}), and the additional ones listed in Sect.~\ref{intro}.

The current census indicates the presence of 159 GCs so far detected in the Milky Way (Sect.~\ref{intro}).
However, a fundamental question is how to define a Galactic GC. The census includes a few dwarf galaxy
nuclei, e.g. $\omega$\,Cen and M\,54 (\citealt{GCProp} and references therein). The recently discovered
SDSS stellar systems in the outer halo (Sect.~\ref{intro}) might be dwarf galaxies. Some clusters not far
from the plane, like Palomar\,1 with a relatively young age (\citealt{Rosenberg98}), might be old open
clusters.

In the case of a GC, FSR\,584 would be the $\rm160^{th}$ detected in the Galaxy. FSR\,584 is remarkably
close to the Sun at $\ds=1.4\pm0.2$\,kpc ($\approx1$\,kpc beyond the Solar circle), thus resulting as
the nearest GC. It is located close to the Galactic plane ($Z\approx20$\,pc), and is moving towards the
plane, according to the PM results (Sect.~\ref{CentralCMD}).

The photometric and structural parameters of FSR\,584 are consistent with a Palomar-like halo GC, i.e. a
low mass GC containing some giants. The TO of FSR\,584 is readily detected in the present near-IR CMDs
(consistent with the proximity), as well as the MS and possibly some blue HB stars. A metallicity of
$\feh\approx-2.0$ is estimated, together with an absorption of $A_V=9.2\pm0.6$. Its Palomar-like nature,
together with the fact that the bulk of its stars require IR photometry due to a relatively high absorption,
can explain why it has so far been overlooked. FSR\,584 is projected in the direction of the molecular cloud
W\,3, and evidence is found that they actually interact, which appears to be the major source of absorption.
Disk shocking and the possible encounter with a massive cloud may have enhanced the dynamical evolution of
FSR\,584, as suggested by a central density excess in the radial density profile. Regardless of the globular
or open cluster nature, the old age of FSR\,584 makes it a promising dynamical laboratory to search for
significant stellar losses and/or disruption related to the potentially damaging environment.

The fact that in less than one year, five new Palomar-like GCs have been identified, AL\,3 in the bulge,
SEGUE\,1, Koposov\,1 and 2, and now possibly FSR\,584 in the halo, suggests that the number of low-mass
GCs may be considerably larger than previously thought.

\begin{acknowledgements}
We thank the anonymous referee for suggestions.
We acknowledge partial support from CNPq and FAPESP (Brazil), and MURST (Italy).
\end{acknowledgements}


\end{document}